\documentclass{kluwer}
\usepackage{epsfig}
\def\kms{\relax \ifmmode {\, \rm km\, s}^{-1}\else \, km\, s$^{-1}$\fi}
\def\ha{\relax \ifmmode {\rm H}\alpha\else H$\alpha$\fi}
\def\hb{\relax \ifmmode {\rm H}\beta\else H$\beta$\fi}
\def\hi{\relax \ifmmode {\rm H{\sc i}}\else H{\sc i}\fi}  
\def\hii{\relax \ifmmode {\rm H{\sc ii}}\else H{\sc ii}\fi}
\def\sii{\relax \ifmmode {\rm S\, {\sc ii}}\else S\, {\sc ii}\fi}
\def\lha{\relax \ifmmode L_{{\rm H}\alpha}\else $L_{{\rm H}\alpha}$\fi}
\def\shi{\relax \ifmmode \sigma_{{\rm HI}}\else $\sigma_{\rm HI}$\fi}  
\def\sh2{\relax \ifmmode \sigma_{{\rm H}_2}\else $\sigma_{{\rm H}_2}$\fi}

\def\Msun{M$_\odot$}

\def\me{$^{-1}$}              %  superscript -1
\def\ref#1{{\hangindent=\parindent \hangafter=1 \par \noindent #1}}
\def\hii{\relax \ifmmode {\rm H\, {\sc ii}}\else H\, {\sc ii}\fi}
\def\ha{\relax \ifmmode {\rm H}\alpha\else H$\alpha$\fi}
\def\kms{\relax \ifmmode {\, \rm km\, s}^{-1}\else \, km\, s$^{-1}$\fi}
\def\Msun{M$_\odot$}
\def\sii{\relax \ifmmode {\rm S\, {\sc ii}}\else S\, {\sc ii}\fi}
\def\deg{\hbox{$^\circ$}}
\def\apj#1,  {{\it ApJ, ~}{\bf#1},  }
\def\apjlett#1,  {{\it ApJL, ~}{\bf#1},  }
\def\apjsupp#1, {{\it ApJS, ~}{\bf#1},  }
\def\aj#1,  {{\it AJ, ~}{\bf#1},  }
\def\astrf#1,  {{\it Astrofizika, ~}{\bf#1},  }
\def\aasupp#1,  {{\it AAS, ~}{\bf#1},  }
\def\aa#1,  {{\it AA, ~}{\bf#1},  }

\def\mnras#1,  {{\it MNRAS, ~}{\bf#1},  }
\def\mmnras#1,  {{\it MemRAS, ~}{\bf#1},  }

\def\annrev#1,  {{\it ARAA, ~}{\bf#1},  }
\def\ass#1,  {{\it Astrophys.\ Space\ Sci.~}{\bf#1},  }
\def\pasp#1,  {{\it PASP, ~}{\bf#1},  }
\def\pasa#1,  {{\it PASA, ~}{\bf#1},  }

\def\prl#1,  {{\it Phys. Rev. Letters, ~}{\bf#1},  }         
\begin{document}

\begin{frontmatter}

\end{frontmatter}
\begin{article}
	\begin{opening}
		\title{The \ha\ emission of the spiral galaxy NGC~7479}
		
		\author{Almudena \surname{Zurita} \email{azurita@ll.iac.es}} 
		
		\author{Maite \surname{Rozas} \email{mrozas@ll.iac.es}}
		
		\author{John E. \surname{Beckman} \email{jeb@ll.iac.es}}
		
		\institute{Instituto de Astrof\'\i sica de Canarias, 38200-La
		Laguna, Tenerife, SPAIN}
		
		\begin{abstract}
		We use the catalogue of \hii\ regions obtained from a high  quality
continuum-subtracted H$\alpha$ image of the grand design spiral galaxy NGC~7479, 
to construct the
luminosity function (LF) for the \hii\  regions (over 1000) of the whole galaxy.
Although its slope is within the published range for spirals of the same morphological type, the
unusually strong star formation along the intense bar of NGC~7479 prompted us to
analyze separately 
the \hii\ regions in the bar and in the disc.
We have calculated the physical properties of a group  of  \hii\ regions in the bar and in the disc
selected for their regular shapes and absence of blending.
We have obtained galaxy-wide relations for the \hii\ region set: diameter distribution function and also the global H$\alpha$ surface density distribution.
As found previously for late-type spirals, the disc LF shows clear double-linear behaviour with a
break at log~L$_{H\alpha}$=38.6 (in erg s$^{-1}$). The bar LF is less regular.
This reflects a physical difference between the bar and the disc in the properties of their 
populations of regions.
		\end{abstract}
	\end{opening}
	
\section{Observations, data reduction and production of the \hii\ region catalogue.}

The observations were made through the TAURUS camera on the 
4.2m William Herschel Telescope
on La Palma. The detector used was an EEV CCD 7 with projected pixel size 
0''.279 $\times$ 0''.279.
Observing conditions were good, with photometric sky and 0''.8 seeing.
After standard reduction routines, we obtained the calibrated H$\alpha$ continuum-subtracted image
(Fig. 1) by subtracting an image through a non-redshifted filter to the image obtained through a 15
\AA\ filter with central wavelength equal to the redshifted H$\alpha$ emission from the galaxy
and  calibrating with observations of standard stars.
\begin{table}[htb]
\caption[ ]{NGC~7479: basic parameters (data from RC3 catalogue, except  P.A. and {\em i}, both 
from Laine \&
Gottesman 1998)}
\begin{tabular}{llllll}
\hline 
{\bf R.A. (2000)}    & 23$^h$ 04$^m$ 57$^s$.1 & {\bf Type}	& SBS5&{\bf {\em i}}    &  51$^{\circ}$\\
{\bf Dec (2000)}     & 12$^{\circ}$ 19$^{'}$ 18$^{''}$ &{\bf  D(Mpc)} & 31.94& {\bf v$_{opt}$} &    2394 km/s \\
{\bf r$_{25}$}       &  2.04$^{'}$ & {\bf B$_T$}     &  11.6 & {\bf P.A. }     &   22$^{\circ}$ \\
\hline
\end{tabular}
\end{table}
As a selection criterion to construct the \hii\ region catalogue we specified that a 
feature must contain at least nine
contiguous pixels, each with an intensity of at least three times the
r.m.s noise level of the local background. 
The r.m.s. noise of the
 background-subtracted H$\alpha$ image is ~15  instrumental counts, which means that
lower limits to the luminosity of the detected \hii\ regions, and to the radius
 of the smallest catalogued regions  are, respectively, log L$_{H\alpha}$ =  37.65  erg s$^{-1}$ and
 $\approx 75$ pc.
\begin{figure*}
\centering\epsfxsize=9.2cm \epsfbox[24 140 530 720]{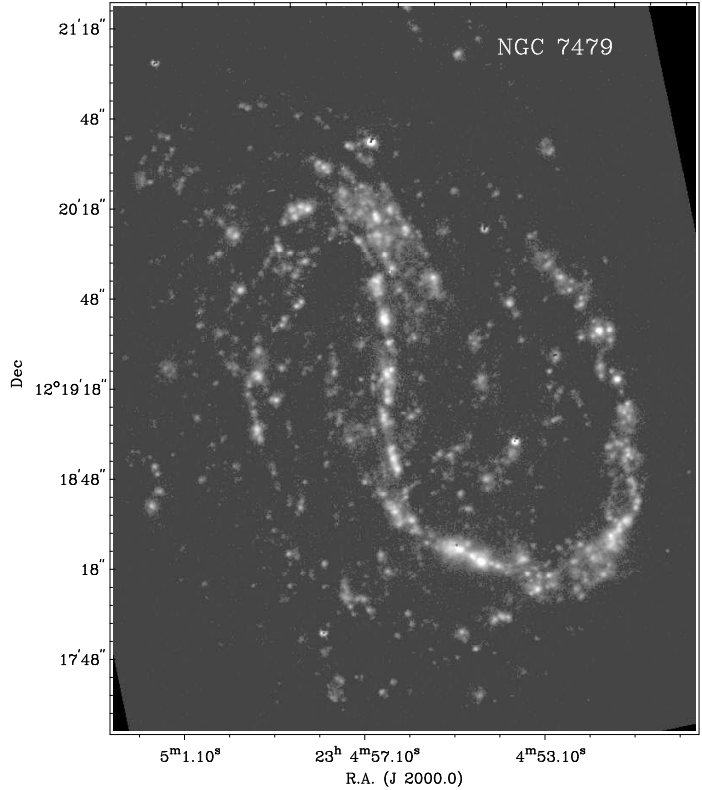}
\caption{Representaction of the H$\alpha$ continuum-subtracted image of NGC~7479}
\end{figure*}
The detection and cataloguing of the \hii\ 
regions were performed using a new program developed by C. Heller.
The program identifies each \hii\ region, measures the position of its
centre, derives the area in pixels  and the flux of each region, integrating all
the pixels belonging to the region and subtracting the local background value.
We catalogued 1009   \hii\ regions in NGC~7479, and
for all the \hii\ regions we determined
equatorial coordinate offsets from the nucleus and deprojected distances
to the centre (in arcsec) using the inclination  and position
angles given by Laine \& Gottesman (1998) ($i=51^{\circ}$, PA=22$^{\circ}$).  In Fig.~2 we
show schematically the positions of the \hii\ regions in the disc of
NGC~7479, on a deprojected RA-dec grid centred on the nucleus of the galaxy.

\begin{figure*}
\centerline{
\epsfig{file=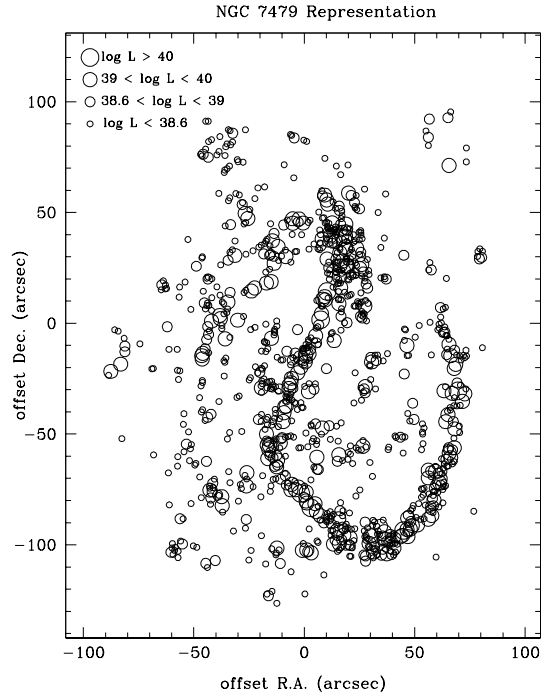,width=10cm}}
\caption{Representation of the positions of the measured \hii\ regions. 
Symbols show ranges of log L. Coordinates of the centre of the image are R.A.=23h 
4m 56.64s Dec=12\deg 19$'$ 22.9$''$ (J2000)}
\end{figure*}

\section{Luminosity functions}

The full count of \hii\ regions seen in Fig. 2  is presented in
differential form,  as a  log-log  plot in bins of 0.15 dex,  in Fig. 3 . The 
luminosity range in \ha\
is limited at the high end by the luminosity of the brightest detected region, 
and at the low end by our criteria for a detection. The limit of
 completeness is   log L$_{H\alpha}$ = 38.0, so that the apparent broad peak in 
the distribution just below
this luminosity is an artefact due to observational selection. 
The best single linear fit of the data to the
points above  log L$_{H\alpha}$ =38, corresponding to a power law
\begin{center}
\vspace{-0.2cm}
 $dN(L)=AL^{\alpha}dL$, 
 \vspace{-0.2cm}
\end{center}
has a gradient  -0.83$\pm$0.06 (i.e. $\alpha$=-1.83$\pm$0.06). We have also carried
out a bi-linear fit,  based on the evidence that at log L $\sim$ 38.8 there is an apparent
 change in slope in the LF and we explained it as a manifestation of a transition in the physical 
properties of the regions, but in those cases the luminosity of the transition was 
always log L$_{Str}\sim$ 38.6 ($\sim$ 0.2 dex lower than in NGC~7479).
As NGC~7479 is a barred galaxy with unusually strong star formation in the bar, 
we tested the hypothesis that the differences in the LF might be due to this intense 
star formation activity under somewhat different physical conditions from those in the disc.
To perform this test   we  constructed separately the LF's for the \hii\ regions of
the bar and of the disc. The results are shown in Fig. 3 and the LF's slopes are in Table II.
In the LF for the 629 \hii\ regions detected in the disc  (Fig. 5) the
change in slope is cleaner than for the total LF and occurs at log~L=38.65$\pm$0.15 (in erg s\me).
The change of slope, accompanied by a slight bump in the LF, has been detected in the seven 
galaxies so far examined in this degree of detail (see e.g. Rozas, Beckman \& Knapen 1996, 
Knapen et al. 1993).
\begin{figure*}[h]
\epsfxsize=11cm \epsfbox[0 150 560 710]{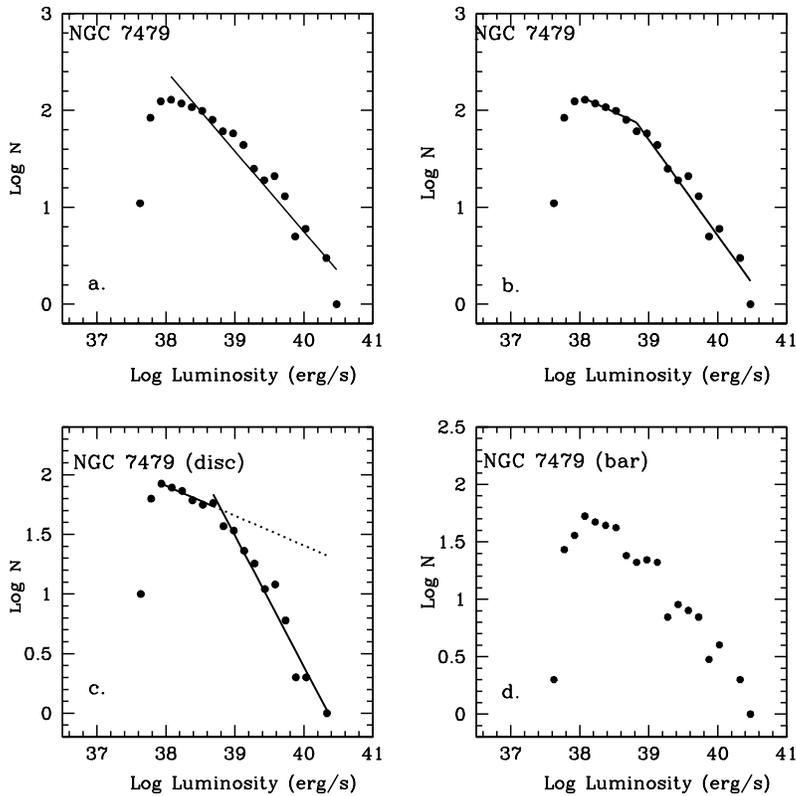}
\caption{Luminosity function for the complete sample of \hii\ regions from the ca\-ta\-logue with the best
linear (a) and bi-linear fit (b), based on the previous experience with late-type spirals. (c) shows the
bi-linear fit for the luminosity function of the \hii\ regions of the disc; the change in  slope is seen at
log L$_{H\alpha}$=38.65. The luminosity function of the bar (d) is clearly less well-behaved than that for the
disc.}
\end{figure*}
 Adding the value for NGC~7479 to the set of values previously measured for other galaxies, 
 we find that the 
rms scatter in L$_{Str}$ in the full set of objects is  0.08 mag. This low scatter can be
explained if the IMF at the high luminosity end of the mass function changes little from
galaxy to galaxy,  {\em i.e.} varies little with metallicity.  Another
necessary condition is that the rate of emission of ionizing photons from a young stellar
cluster rises more rapidly than the mass of its placental cloud, a condition which we have
examined observationally (Beckman et al. 1999), and shown to hold. 
If we look at Fig. 3, where the LF for the 380 \hii\ regions of the bar is
presented, we can see clearly that the irregularity found in the total LF is due to the
incorporation of  the \hii\ regions of the bar in the total LF. 
Clearly star formation conditions in the bar differ from those in the disc.
\begin{table}[htb]
\caption[ ]{Luminosity limits, LF slopes and correlation coefficients for the two luminosity ranges (in erg s$^{-1}$)
for the LF for the complete sample of \hii\ regions from the catalogue and for the LF of the
\hii\ regions of the disc.}
\vspace{-0.2cm}
\begin{tabular}{ccccc}
\hline 
& & {\bf range log L} & {\bf $\alpha$} & {\bf r}\\
\hline 
{\bf Total LF} & single linear fit & 38.1$\geq$log L$\leq$40.5 & -1.83$\pm$0.06 & 0.969 \\
               & bi-linear fit      & 38.1$\geq$log L$\leq$38.8 & -1.33$\pm$0.04 &  0.979 \\
               &                   & 38.8$\geq$log L$\leq$40.5 & -1.99$\pm$0.07 &  0.978 \\
\hline	 
{\bf Disc LF}& single linear fit & 37.9$\geq$log L$\leq$40.4 & -1.82$\pm$0.07  &0.958 \\
             & bi-linear fit      & 37.9$\geq$log L$\leq$38.6 & -1.25$\pm$0.04 &  0.956\\
             &                   & 38.6$\geq$log L$\leq$40.4 & -2.10$\pm$0.07&  0.980 \\
\hline
\end{tabular}
\end{table}
\vspace{-0.6cm}

\section{Other Statistical properties.}

\begin{itemize}
\item {\bf The integral diameter distribution}  (number of regions with
 diameters greater than a given 
value as a function of diameter) is given in Fig. 4a From our own and other published 
studies  it has been found that
the diameter distribution of the \hii\ regions can be well fitted by an exponential of form 
%\begin{center}
%\vspace{-0.2cm}
$N(>D)= N_{\rm o}\, exp(-D/D_{\rm o})$ 
%\end{center}
%\vspace{-0.2cm}
where $D_{\rm o}$ is a characteristic diameter, and $N_{\rm o}$ is an 
(extrapolated) characteristic value for 
the total number of regions. From the observations represented in Fig. 4a,   we
obtain the values $N_{\rm o}$=5000$\pm$300 and $D_{\rm o}$= 110$\pm$2 pc (with 
H$_{\rm o}$=75 \kms Mpc$^{-1}$).
 $D_{\rm o}$ has a  value predicted for a galaxy of its measured absolute 
luminosity according to the observational plot of 
$D_{\rm o}$ versus luminosity (Hodge 1987).
% The value 
%for N$_{\rm o}$ is within the range found for previously observed late type spirals.
 
\item {\bf The flux density distribution} is  shown in Fig. 4b. It was found by  dividing
 the disc into rings using the values of P.A. and {\em i} of the galaxy 
 (Table I).
 There is a clear trend to lower flux,   with 
increasing radius,  which clearly reflects the standard radial decline in surface
density of each major component of this disc galaxy. This underlying radial flux density
 can be fitted by an exponential of form $f(r) = B \, exp({-{\rm r}/{\rm h}_{H\alpha}})$
from which we derive a value for the \ha\ scale length, {${\rm h}_{H\alpha}$= 
(2.4$\pm$0.3) kpc}.
\end{itemize}
\begin{figure}
\centerline{
\epsfig{file=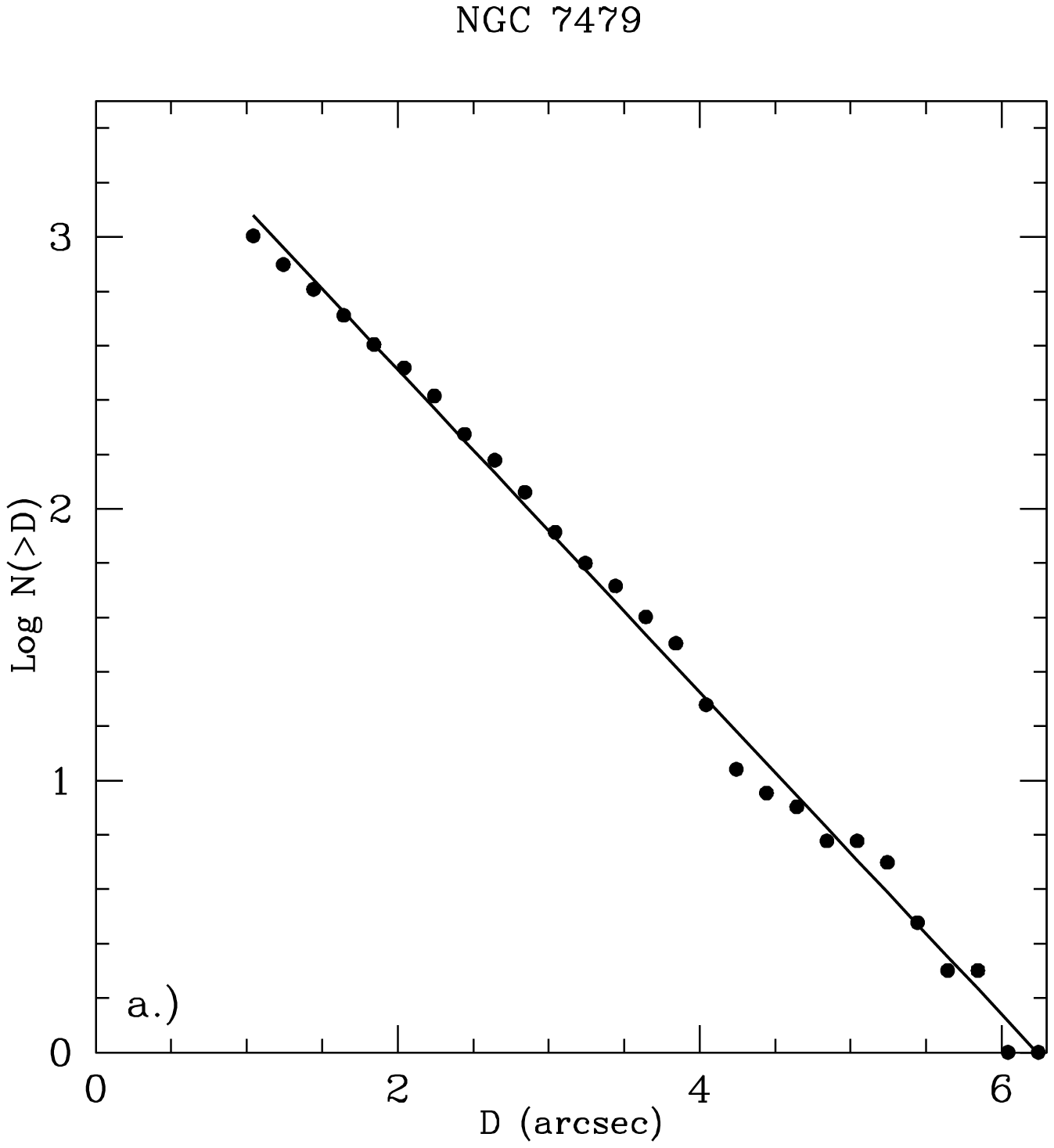,width=9.0cm}\qquad
\hspace{-2.2cm}
\epsfig{file=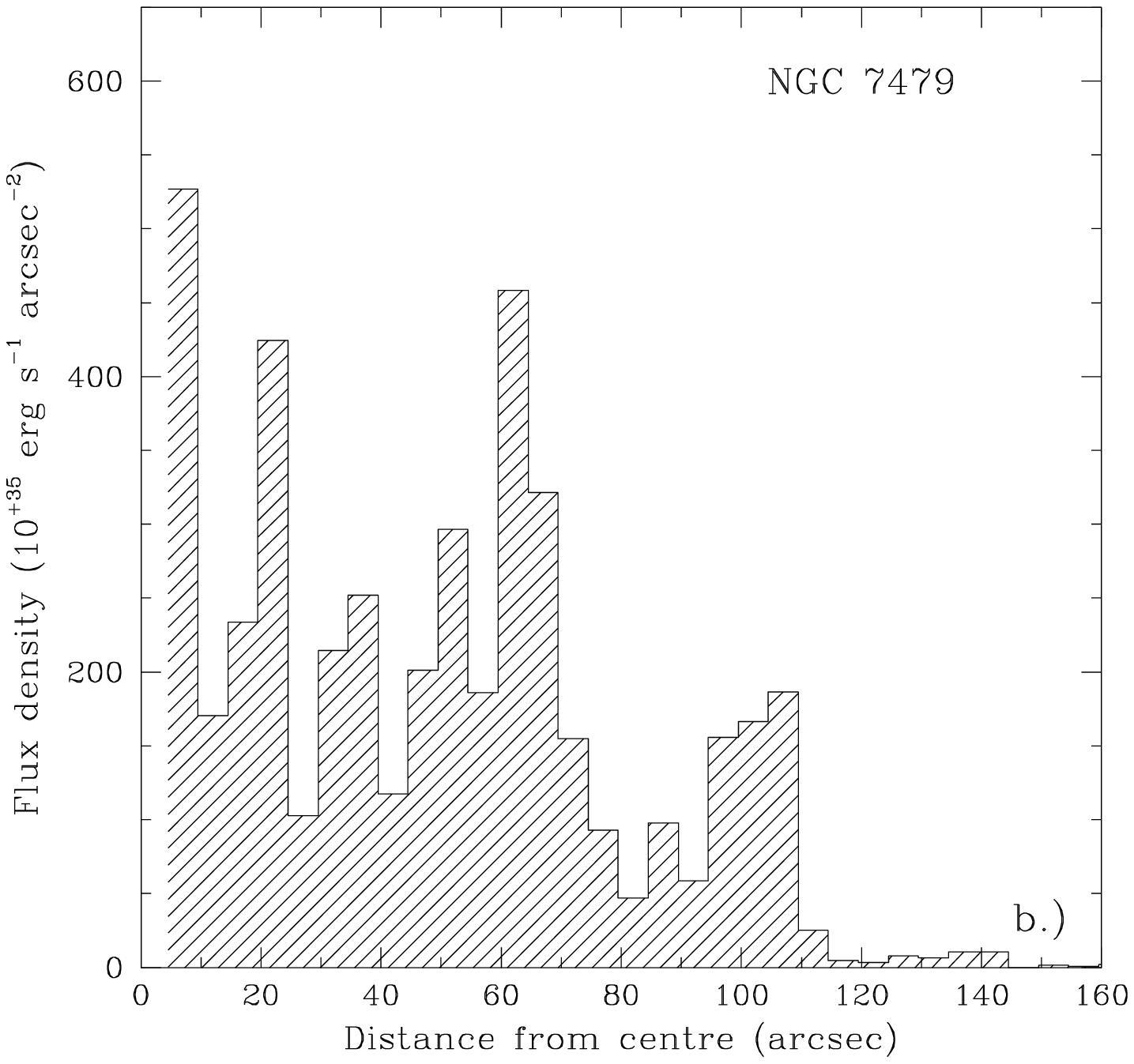,width=9.0cm}
\vspace{-1.8cm}
\caption{a.)Integral logarithmic diameter distribution function of all \hii\ regions of NGC~7479.
  The straight line indicates the best fit. b.)Flux density distribution of all the \hii\ regions of NGC~7479 as a function of the
 deprojected distance to the centre of the galaxy.}}
\end{figure}

\section{Physical properties.}

Using a distance to NGC~7479 of 31.92 Mpc we employed the standard theoretical formula (Spitzer  1978)
%\begin{center}
%$\frac {\rm L} {4\pi r^2}$=$h \nu_{H\alpha}\alpha_{H\alpha} \frac {n_p} {n_e} 2.46\times10^{17} Em $
%\end{center}
relating the
surface brightness of an \hii\ 
region with its emission measure (Em),  (assuming case B recombination and T$\sim$10$^4$ K)  to calculate the Em of the  \hii\ regions.
 We performed this for a total of 39 regions, (22 in the disc and the rest in the bar), covering the
full range of observed radii and chosen as isolated to minimize the uncertainties in calculating their 
luminosities due to the overlapping.  Results are shown in Fig. 5a. 

In Fig. 5b. we show the rms electron density (derived from the Em, $<~N_{e}>_{rms}=\sqrt{Em/r}$) 
plotted  
%against radius of
%the \hii\ region and against luminosity, for the same \hii\ regions, in Fig. 5c. 
against radius of the \hii\ region . 
The general ranges and behaviour of Em and $<N_{e}>_{rms}$ for NGC~7479 agree well with those
found by Kennicutt 1984 and by ourselves (Rozas, Knapen \& Beckman  1996) for extragalactic 
\hii\ regions. Due to observational selection,  these tend to be more
luminous and larger than Galactic regions.

We also computed the filling factors, using $\delta$=$(<N_{e}>_{rms}/N_{e})^2$. 
The  implicit
model is that an \hii\ region is internally clumpy, so that the observed flux
comes from a high density component, which occupies a fraction $\delta$ (filling factor) of the
total volume; the rest of the volume is filled with low density gas which
makes a negligible contribution to the observed emission line strengths.

To calculate $\delta$, we need
to know the in situ $N_{e}$ for each region. We have not measured these values for NGC~7479, 
but have used a ``canonical'' mean value of 135 cm$^{-3}$ obtained by Zaritzky
et al. 1994 for 42 \hii\ regions in a large sample of galaxies. The 
filling factors for the regions range from 4.3$\times10^{-4}$ to 1.5$\times10^{-3}$,
 a range which coincides well with those found for 4 galaxies in  Rozas et al. 1996b.  
 
Values of $<N_{e}>_{rms}$ can also be used to estimate the mass of ionized
gas, which range from 3000\Msun\ to 1.5$\times10^{6}$\Msun. These physical properties, the emission of
Lyman cotinuum photons and the equivalent number of O5V stars required to supply the luminosity of the
regions are given in a more detailed paper (Rozas, Zurita \& Beckman 1998).

\begin{figure*}[h]
\center
%\hspace{-2.7cm}
\epsfxsize=8.6cm \epsfbox[56 240 546 466]{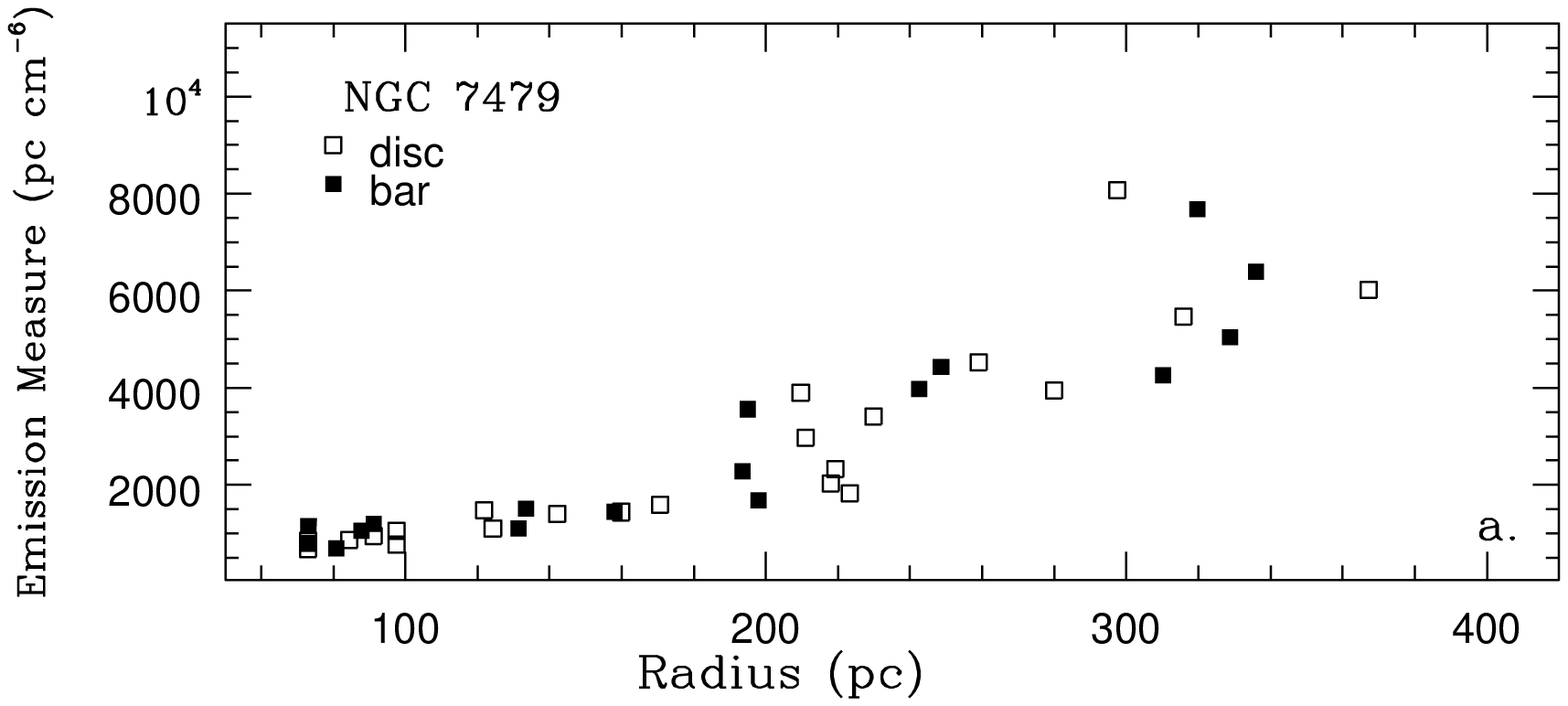}
%\hspace{-0.1cm}
\epsfxsize=8.6cm \epsfbox[56 240 546 466]{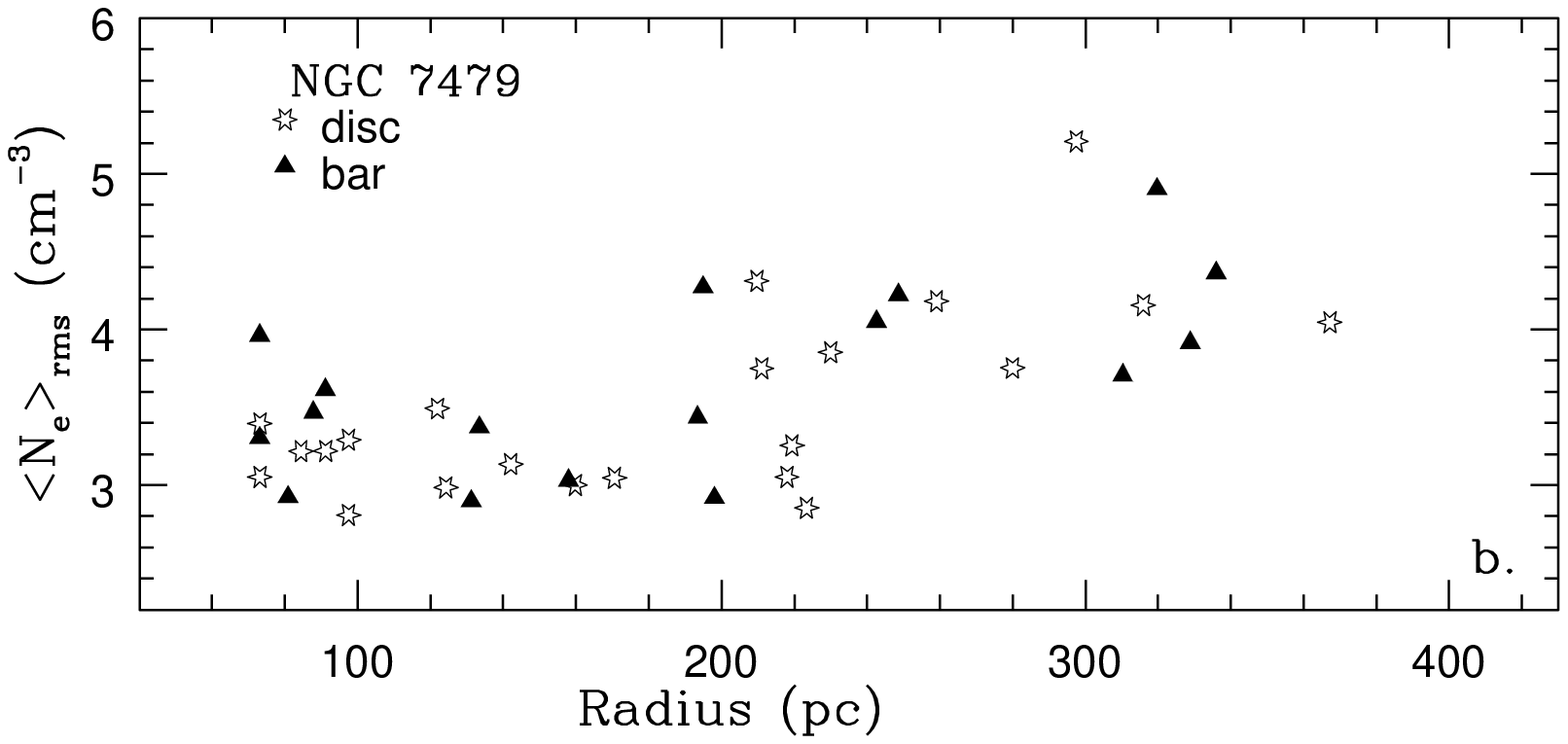}
\center
\caption{Emission measure versus radius (a.), {\em rms} electron density versus radius (b.) for the 39 
selected \hii\ regions in NGC~7479.
%and {\em rms} electron density versus log L (c.) for the selected \hii\ regions.
}
\end{figure*}

\section{Conclusions}
The main results of the present work, where we analyze the statistics and properties 
of \hii\ regions in NGC~7479 are summarized below.

\begin{itemize}
%\vspace{-0.1cm}
\item Using a high quality continuum-subtracted H$\alpha$ image of the
grand-design spiral NGC~7479,  we have
catalogued  1009 \hii\ regions.

%\item The total flux emitted in H$\alpha$ by NGC~7479 is
%L$_{H\alpha}$(total)=(1.3$\pm$0.2)$\times$10$^{42}$ erg/s which indicates 
%that NGC~7479 is a high luminosity galaxy in H$\alpha$.
\vspace{-0.2cm}
\item  The slope of the LF  agrees broadly
with slopes for other galaxies of comparable morphological types.

\vspace{-0.2cm}
\item We have found a change in slope in the LF of the \hii\ regions of NGC~7479 that occurs at a
luminosity slightly higher ($\simeq$ 0.2 dex) than that found in other galaxies of the same
morphological type. Due to the intense star formation in the bar of NGC~7479, we decided to construct separately
the LF's for the \hii\ regions of the bar and of the disc, finding:
\begin{enumerate}
\vspace{-0.2cm}
\item the LF of the disc is
in no way different from that found in previous papers for galaxies of the same morphological
type,

\item  The anomaly in the global LF is thus due to different star formation conditions 
in the bar. This must be due to the effects of gas dynamical parameters on the stellar IMF, and the  
 physical conditions in the clouds.
 \end{enumerate}
\vspace{-0.2cm}
\item The integrated distribution function of the \hii\ region diameters can be well
fitted by a exponential function. 

\vspace{-0.2cm}
\item The densities, filling factors and masses derived
from the luminosities and sizes of a selected set of representative
regions, through the range of observed luminosities for NGC~7479,
are in agreement with those found in the previous literature on
extragalactic \hii\ regions. 
%ionization indices
%\item The physical properties encountered cover
%a wide range, but all have moderately low electron densities, always much
%less than 10 cm$^{-3}$, of the order of mean interstellar number
%densities in galaxies. 
\end{itemize}

\end{article}

\end{document}